\documentclass{iaus}
\usepackage{graphicx}
\usepackage{natbib}

\def\simless{\mathbin{\lower 3pt\hbox
   {$\rlap{\raise 5pt\hbox{$\char'074$}}\mathchar"7218$}}}
\def\simgreat{\mathbin{\lower 3pt\hbox
   {$\rlap{\raise 5pt\hbox{$\char'076$}}\mathchar"7218$}}}

\begin{document}

\title{The Chemistry of the Early Universe}
\author[Glover et~al.]{S.~C.~O. Glover}

\affiliation{Institut f\"ur Theoretische Astrophysik, Zentrum f\"ur Astronomie der Universit\"at Heidelberg, Albert-Ueberle-Str.\ 2, 69120 Heidelberg, Germany}

\pubyear{2011}
\jname{The Molecular Universe}
\editors{J.\ Cernicharo \& R. Bachiller}

\maketitle

\begin{abstract}
The chemistry of the early Universe is a fascinating field of study. Even in the absence of any elements heavier than lithium,
a surprising degree of chemical complexity proves to be possible, giving the topic considerable interest in its own right.
In addition, the fact that molecular hydrogen plays a key role in the formation of the first stars and galaxies means that
if we want to understand the formation of these objects, we must first develop a good understanding of the chemical
evolution of the gas. In this review, I first give a brief introduction to the chemistry occurring in the gas prior to the formation 
of the first stars and galaxies, and then go on to discuss in more detail the main chemical processes occurring during the
gravitational collapse of gas from intergalactic to protostellar densities, and how these processes influence the final outcome
of the collapse.
\end{abstract}

\firstsection

\section{Introduction}
\label{sec:intro}
Chemistry in the early Universe has proved to be an interesting topic of study for several reasons. Before the first stars and 
galaxies formed, the elemental composition of the gas was very simple -- it consisted only of hydrogen and helium (and
their isotopes) plus a tiny trace of lithium -- and yet this is sufficient to give rise to a surprising degree of chemical complexity.
In addition, the interaction between the gas and the cosmic microwave background (CMB) plays an important role in 
regulating the chemistry of the gas, and moreover one which may lead to observable changes in the CMB 
\citep[see e.g.][]{maoli94,dub08,sch08}. Furthermore, understanding the chemical evolution of the gas at these early epochs is vital 
if one wants to understand the formation of the first stars and galaxies, given the crucial role played by molecular hydrogen 
(H$_{2}$) in this process. 

There is a wide range of topics that one could discuss in a review of the chemistry of the early universe, ranging from the
physics and chemistry of the recombination epoch at redshift $z > 800$ to the reionization of the Universe at $z < 15$.
However, to keep this review to a manageable length, I will focus on only a few key issues, and will not discuss either
the recombination epoch or the process of cosmological reionization. Both of these topics are treated in detail elsewhere
(see e.g.\ \citet{fendt09} and references therein for a lengthy discussion of the physics of the recombination epoch,
or the recent review by \citet{mw10} for a discussion of the epoch of reionization).

In the next Section,  I review what we know about the chemical evolution of 
the IGM from the end of the recombination epoch until the time at which the earliest protogalaxies form. I follow this up in
Section~\ref{proto} by discussing the chemical evolution of the gas within these early protogalaxies. I show how an understanding 
of the gas chemistry can allow us to determine the mass scale of the first star-forming minihalos, outline the main chemical 
changes that occur in the gas during its collapse to protostellar densities, and highlight the important role played by molecular
hydrogen throughout this process.

\section{Pregalactic chemistry}
\label{pre}
\subsection{The chemical composition of the gas at the end of the recombination epoch}
At the end of the recombination epoch, at a redshift $z \sim 800$, the chemical makeup of the gas in the Universe -- what we
might refer to as pregalactic gas -- was very simple. The main constituent of the gas was neutral atomic hydrogen, accounting 
for roughly 75\% of the total mass. A small amount of hydrogen remained ionized, with the ratio by number of ionized to neutral 
hydrogen being roughly ${\rm H^{+}}/{\rm H} \simeq 2.5 \times 10^{-3}$ (\citealt{wms08}; note that this value continues to decline
at lower redshift, decreasing by about an order of magnitude by the time we reach $z = 100$). Neutral atomic helium (primarily 
$^{4}$He, but also including a small fraction of $^{3}$He) accounted for most of the remaining mass, but in this case the abundance 
of singly or doubly ionized helium was very close to zero, as helium recombination comes to an end at a significantly higher redshift 
than hydrogen recombination \citep{wms08}. Deuterium was also present, with a fractional abundance relative to hydrogen 
${\rm D}/{\rm H} \simeq 2.5 \times 10^{-5}$ \citep{cyb08}, and a ratio of D$^{+}$ to D that was essentially the same as the ratio of 
ionized to neutral hydrogen. Finally, there was a trace amount of lithium (${\rm Li/H} \simeq 5 \times 10^{-10}$; \citealt{cyb08}), the 
majority of which was in the form of Li$^{+}$. The fractional abundances of all other ionized or molecular species were smaller 
than $10^{-12}$ at this point \citep{ah10}.

\subsection{The chemical evolution of the IGM}
Starting from this simple beginning, many other chemical species were produced as the Universe evolved. Detailed chemical models 
for the evolution of pregalactic gas have been constructed by a number of authors \citep[see e.g.][]{sld96,sld98,gp98,lsd02,ps02,sch08},
but in this discussion I will focus on a few general principles that can help us to understand the chemical evolution of the gas at these
early times.

As a starting point, it is useful to draw a distinction between what I will call ``primary'' species, i.e.\  species that are formed
directly from chemical reactions amongst the main chemical constituents of the gas (H, H$^{+}$, D, D$^{+}$, He and Li$^{+}$),
and ``secondary'' species that are formed mainly from reactions involving one or more of the primary species. Good examples of 
primary species are H$^{-}$, formed via the reaction
\begin{equation}
{\rm H} + {\rm e^{-}} \rightarrow {\rm H^{-}} + \gamma,  \label{ra}
\end{equation}
or H$_{2}^{+}$, formed via the reaction 
\begin{equation}
{\rm H} + {\rm H^{+}} \rightarrow {\rm H_{2}^{+}} + \gamma.
\end{equation}
The prime example of a secondary species is H$_{2}$, as direct formation of H$_{2}$ via the reaction
\begin{equation}
{\rm H} + {\rm H} \rightarrow {\rm H_{2}} + \gamma
\end{equation}
occurs at a negligible rate \citep{gs63}, and most H$_{2}$ in metal-free gas forms via the reactions
\begin{equation}
{\rm H^{-}} + {\rm H} \rightarrow {\rm H_{2}} + {\rm e^{-}} \label{ad}
\end{equation}
and
\begin{equation}
{\rm H_{2}^{+}} + {\rm H} \rightarrow {\rm H_{2}} + {\rm H^{+}}.
\end{equation}

The evolution of the fractional abundance of a primary species X is largely governed by the relative
sizes of three important timescales: the formation time, defined as
\begin{equation}
 t_{\rm form, X} = \frac{n_{\rm X}}{R_{\rm form, X}}
\end{equation}
where $n_{\rm X}$ is the number density of species X and $R_{\rm form, X}$ is the formation rate
per unit volume of X; the photodissociation time,
\begin{equation}
t_{\rm pd, X} = \frac{1}{R_{\rm pd, X}},
\end{equation}
where $R_{\rm pd, X}$ is the photodissociation rate of species X; and the Hubble time
\begin{equation}
t_{\rm H} = \frac{1}{H(z)},
\end{equation}
where $H(z)$ is the Hubble constant.

At high redshift, photodissociation of most molecules and molecular ions by the CMB occurs rapidly, 
and the photodissociation time is small. Similarly, the high gas density means that the formation timescale
of most species is also small. In this regime, both timescales are typically much shorter than the Hubble time, and
hence the abundance of species $X$ evolves until it reaches an equilibrium value set by the balance
between formation and photodissociation (in which case $t_{\rm form, X} = t_{\rm pd, X}$). Since photodissociation
is very effective at high redshift, this equilibrium value is generally very small. 

As we move to lower redshifts, the cosmological background density decreases, since $\rho(z) \propto (1+z)^{3}$,
and hence the formation time of any given primary species tends to increase. However, the energy density of
the CMB decreases even more rapidly, $\rho_{\rm rad} \propto (1+z)^{4}$, and the CMB temperature decreases
as $T_{\rm CMB} \propto (1+z)$. What this means is that we eventually reach a redshift at which the number 
density of photons capable of photodissociating species X starts to fall off exponentially, as the threshold energy
for photodissociation starts to fall within the Wien tail of the CMB spectrum.\footnote{This simple picture is complicated
somewhat by the presence of non-thermal photons produced during hydrogen recombination that
slightly distort the spectral shape of the CMB \citep{sh05,hp06}, but even in this
case the general principle is the same.} Once we reach this point, $t_{\rm pd}$ rapidly increases, and we are left 
with a situation in which $t_{\rm form}$ is by far the shortest of the three timescales. The abundance of species X
therefore increases rapidly until $t_{\rm form, X} \simeq t_{\rm H}$. Beyond this point, the abundance of X does
not increase significantly, as the time required would be longer than the age of the Universe, and we speak of
the abundance of the species ``freezing out'' at some asymptotic value.

The evolution with redshift of secondary species such as H$_{2}$ is harder to generalize in this way, as it often 
depends on the behaviour of more than one primary species, each of which passes through the three stages noted
above at a different redshift. For example, H$_{2}$ formation via the H$_{2}^{+}$ ion becomes possible at a 
redshift $z \sim 400$, but H$_{2}$ formation via the H$^{-}$ ion is suppressed until $z \sim 100$, owing to the
fact that H$^{-}$ has a much lower binding energy than H$_{2}^{+}$ (0.755~eV for the former, 2.65~eV for the
latter), and hence can be destroyed by photons with significantly smaller energies. The H$_{2}$ abundance in
the pregalactic gas therefore passes through two periods of relatively rapid growth, one occurring around
$z \sim 400$ and the second around $z \sim 100$. Following this second period of growth, the fractional
abundance of H$_{2}$ is of the order of $10^{-6}$, and H$_{2}$ is the most abundant molecule in the gas.
However, this is still far smaller than the typical amount of H$_{2}$ produced within the earliest star-forming
protogalaxies, and hence although the chemical evolution of the pregalactic gas is interesting, it is of limited
influence: almost all of the chemistry affecting the formation of the first stars and galaxies occurs within the
protogalaxies themselves, as I explore in more detail in the next section.

\section{Protogalactic chemistry}
\label{proto}
As we move to lower redshifts, the distribution of matter in the Universe becomes increasingly inhomogeneous.
The small perturbations in the dark matter distribution that were present during the recombination epoch grow
with time, evolving initially as $\delta \propto t^{2/3}$, but eventually entering the regime in which their evolution
becomes non-linear. These perturbations undergo runaway gravitational collapse and virialization, forming 
dense, gravitationally bound structures that are often referred to as dark matter halos. In the $\Lambda$CDM
model, the earliest halos to form do so on the smallest scales, with larger, more massive halos forming at later
times.

Very little gas accumulates within the earliest halos, as it is prevented from undergoing gravitational collapse
by its own thermal pressure, and also by the significant relative velocity that exists between the gas and the
dark matter \citep{th10}. However, once the dark matter halos become sufficiently massive, their gravitational attraction
dominates over these effects and substantial quantities of gas flow into them, being heated by adiabatic
compression and shocks in the process. In regions where the relative velocity between gas and dark matter
is very small, this occurs once the mass of the dark matter halo exceeds $M_{\rm min} \sim 2 \times 10^{4}
\: {\rm M_{\odot}}$ \citep{nb07}, but in more
typical regions where there is a relative velocity between gas and dark matter of several kilometers per second,
the minimum halo mass is higher, $M_{\rm min} \sim 10^{5} \: {\rm M_{\odot}}$ \citep{tbh10}. 
Low-mass halos with masses close to
$M_{\rm min}$ -- often referred to as ``minihalos'' to distinguish them from their considerably more massive,
low-redshift brethren -- are therefore the first objects that have the potential to form stars. Whether or not the
gas within them can actually form stars depends on how rapidly it can cool. This in turn depends on the
chemical evolution of the gas within the minihalos, a subject that I address below.

\subsection{H$_{2}$ formation and the onset of cooling}
As gas falls into a typical minihalo, it is heated up to a temperature close to the virial temperature of the minihalo
\begin{equation}
T_{\rm vir} \simeq 1200 \left(\frac{M}{10^{6} h^{-1} \: {\rm M_{\odot}}}\right)^{2/3}
 \left(\frac{1+z}{10} \right) \: {\rm K},  \label{tvir}
\end{equation}
(where this expression assumes that the gas is predominantly atomic hydrogen, and that $z \geq 10$; see \citealt{bl01}
for a more general expression). In the post-shock gas, the electron fraction decreases due to radiative recombination, 
but at the same time H$_{2}$ forms, primarily via reactions~\ref{ra} and \ref{ad}.
The amount of H$_{2}$ that forms in the gas can be estimated using a simple toy model for the chemistry first introduced
by \citet{teg97}. We start by assuming that radiative recombination is the only process affecting
the electron abundance, and writing the rate of change of the electron number density as
\begin{equation}
\frac{{\rm d}n_{\rm e}}{{\rm d}t} = - k_{\rm rec} n_{\rm e} n_{\rm H^{+}},
\end{equation}
where $n_{\rm e}$ is the number density of electrons, $n_{\rm H^{+}}$ is the number
density of protons, and $k_{\rm rec}$ is the recombination coefficient. If we 
assume that ionized hydrogen is the only source of free electrons, implying that
$n_{\rm e} = n_{\rm H^{+}}$, and that the temperature remains roughly constant during
the evolution of the gas, then we can solve for the time evolution of the electron
fraction:
\begin{equation}
x = \frac{x_{0}}{1 + k_{\rm rec} n t x_{0}},
\end{equation}
where $x \equiv n_{\rm e} / n$, $n$ is the number density of hydrogen nuclei, and
$x_{0}$ is the initial value of $x$. If we now assume that all of the H$_{2}$ forms via
the pair of reactions given above, and that the only process competing with reaction~\ref{ad} 
for the H$^{-}$ is mutual neutralization
\begin{equation}
{\rm H^{+}} + {\rm H^{-}} \rightarrow {\rm H + H}, \label{mn}
\end{equation}
then we can write the time evolution of the H$_{2}$ fraction, 
$x_{\rm H_{2}} \equiv n_{\rm H_{2}} / n$, as
\begin{equation}
\frac{{\rm d}x_{\rm H_{2}}}{{\rm d}t} =  k_{\rm ra} x n_{\rm H}  \, p_{\rm ad}, \label{xh2}
\end{equation}
where $k_{\rm ra}$ is the rate coefficient of reaction~\ref{ra}, and $p_{\rm ad}$ is the probability 
that any given H$^{-}$ ion will be destroyed by associative detachment (reaction~\ref{ad}) rather than by 
mutual neutralization. Given our assumptions above, this can be written as
\begin{equation}
p_{\rm ad} = \frac{k_{\rm{ad}}n_{\rm H}}{k_{\rm{ad}} n_{\rm H} + k_{\rm{mn}} n_{\rm H^{+}}},
\end{equation}
where $k_{\rm{ad}}$ is the rate coefficient for reaction~\ref{ad} and $k_{\rm{mn}}$
is the rate coefficient for reaction~\ref{mn}. If we again assume that $n_{\rm e} =
n_{\rm H^{+}}$, and in addition assume that the H$_{2}$ fraction and the fractional
ionization are small, so that $n_{\rm H} \simeq n$, then the expression for $p_{\rm ad}$ can be simplified to
\begin{equation}
p_{\rm ad} = \left(1 + \frac{k_{\rm{mn}}}{k_{\rm{ad}}} x \right)^{-1}.
\end{equation}
Substituting this into Equation~\ref{xh2}, we obtain
\begin{equation}
\frac{{\rm d}x_{\rm H_{2}}}{{\rm d}t} = k_{\rm{ra}} x n_{\rm H} \left(1 + \frac{k_{\rm{mn}}}{k_{\rm{ad}}} x \right)^{-1}.
\end{equation}
We can solve for the time evolution of the H$_{2}$ fraction
\begin{equation}
x_{\rm H_{2}}  = \frac{k_{\rm{ra}}}{k_{\rm rec}} \ln \left(\frac{1 +  x_{0} k_{\rm{mn}} / k_{\rm{ad}} + t / t_{\rm rec}}
 {1 + x_{0} k_{\rm{mn}} / k_{\rm{ad}}} \right), \label{xh2_eq}
\end{equation}
where $t_{\rm rec} = (k_{\rm rec} n x_{0})^{-1} $ is the recombination time, and 
where we have once again assumed that $n_{\rm H} \sim n$. If the initial fraction ionization is small, so
that $x_{0} \ll k_{\rm ad} / k_{\rm mn}$, then we can write this expression in the simpler form
\begin{equation}
x_{\rm H_{2}} =  \frac{k_{\rm{ra}}}{k_{\rm rec}} \ln \left(1 + t / t_{\rm rec} \right).
\end{equation}

This analysis tells us two important things. First, it shows us that the H$_{2}$ fraction evolves only logarithmically with
time: most of the H$_{2}$ forms within the first few recombination times, with H$_{2}$ formation at later times being
strongly suppressed owing to the loss of the free electrons from the gas due to recombination. Second, it shows us that the
amount of H$_{2}$ that forms depends on the ratio of the rate coefficients for H$^{-}$ formation (reaction~\ref{ra}) and
H$^{+}$ recombination,  $k_{\rm{ra}} / k_{\rm rec}$. The value of this ratio is strongly temperature dependent, and can
be written approximately as
\begin{equation}
\frac{k_{\rm{ra}}}{k_{\rm rec}} \simeq 10^{-8} T^{1.5}.
\end{equation}
The amount of H$_{2}$ that forms in the gas therefore increases significantly with increasing temperature. 
As the H$_{2}$ cooling rate is also a strongly increasing function of temperature \citep[see e.g.][]{flower00}, one
finds that there is a reasonably sharp divide occurring at a critical virial temperature $T_{\rm crit} \sim
1000 \: {\rm K}$ between minihalos that contain gas that can cool rapidly and form stars, and minihalos
that contain gas that barely cools over a Hubble time \citep[see e.g.][]{teg97,yahs03}. If we convert this critical 
virial temperature into a corresponding critical minihalo mass using Equation~\ref{tvir}, we find that
\begin{equation}
M_{\rm crit} \simeq 7.6 \times 10^{5} h^{-1} \left(\frac{1+z}{10} \right)^{-3/2}  \: {\rm M_{\odot}}. \label{mcrit}
\end{equation}
At $z > 40$, this value is smaller than the minimum mass $M_{\rm min} \sim 10^{5} \: {\rm M_{\odot}}$ required to 
overcome the effects of the streaming of the gas relative to the dark matter, which therefore sets the minimum mass 
scale for a star-forming minihalo. At $z < 40$, however, $M_{\rm crit} > M_{\rm min}$, and it is the amount of H$_{2}$
that can form within the gas that is responsible for setting the minimum mass scale. One implication of this result 
is that there will be a population of small minihalos that forming at $z < 40$ that contain significant gas fractions,
but that do not form stars, because their gas is unable to cool in less than a Hubble time. These small starless minihalos 
may be important sinks for ionizing photons during the epoch of reionization \citep{ham01}. 

\subsection{The initial collapse phase}
In minihalos where H$_{2}$ cooling is efficient, the gas cools and collapses under its own self-gravity, collecting at the
centre of the minihalo. The evolution of the gas at this stage depends on the amount of H$_{2}$ it is able to form, which
in turn depends upon the initial ionization state of the gas.

During the formation of the very first Population III stars, the initial fractional ionization of the gas is the same as the residual 
ionization in the intergalactic medium, i.e.\ $x_{0} \sim 2 \times 10^{-4}$ at $z \sim 30$. In this case, the amount of H$_{2}$ 
that forms in the gas is typically enough to cool it to a temperature of $T \sim 200 \: {\rm K}$ but not below. At this temperature, 
deuterated hydrogen, HD, forms easily from H$_{2}$ via
\begin{equation}
{\rm H_{2} + D^{+}} \rightarrow {\rm HD + H^{+}},
\end{equation}
but its destruction via the inverse reaction
\begin{equation}
{\rm HD + H^{+}} \rightarrow {\rm H_{2} + D^{+}}
\end{equation}
is starting to become ineffective, owing to the fact that this reaction is endothermic by around 0.04~eV. At $T = 200$~K,
chemical fractionation therefore boosts the HD/H$_{2}$ ratio by a factor of 20 compared to the cosmic deuterium-to-hydrogen
ratio. This boost to the HD abundance, together with the fact that HD is a much more effective coolant than H$_{2}$ at low 
temperatures \citep[see e.g.][]{flower00}, allows HD to become an important coolant at 200~K, but detailed studies have 
shown that the extra cooling provided by HD is not enough to reduce the temperature significantly below 
200~K \citep{bcl02}. Therefore,
H$_{2}$ continues to dominate the cooling and control the further evolution of the gas. In this scenario, the collapse of the 
gas is greatly slowed once its density reaches a value of around $10^{4} \: {\rm cm^{-3}}$, corresponding to the critical density
$n_{\rm crit}$, at which the rotational and vibrational level populations of H$_{2}$ approach their local thermodynamic equilibrium 
(LTE) values. At densities higher than this critical density, the H$_{2}$ cooling rate per unit volume scales 
only linearly with $n$ (compared to a quadratic dependence, $\Lambda_{\rm H_{2}} \propto n^{2}$ at lower densities),
while processes such as compressional heating continue to increase more rapidly with $n$. As a result, the gas 
temperature begins to increase once the density exceeds $n_{\rm crit}$. 

Gas reaching this point in the collapse enters what \citet{bcl02} term a ``loitering'' phase, during which cold gas 
accumulates in the centre of the halo but
only slowly increases its density. This loitering phase ends once the mass of cold gas that has accumulated exceeds
the local value of the Bonnor-Ebert mass  \citep{bonnor56,ebert55}, given in this case by \citep{abn02}
\begin{equation}
M_{\rm BE}  \simeq  40 T^{3/2} n^{-1/2} \: {\rm M_{\odot}},
\end{equation}
which for $n \sim 10^{4} \: {\rm cm^{-3}}$ and $T \sim 200$~K yields $M_{\rm BE} \sim 1000 \: {\rm M_{\odot}}$.
Once the mass of cold gas exceeds $M_{\rm BE}$, its collapse speeds up again, and becomes largely decoupled
from the larger-scale behaviour of the gas. The next notable event to occur in the gas is the onset of three-body 
H$_{2}$ formation, which is discussed in Section~\ref{threeb} below.

If the initial fractional ionization of the gas is significantly higher than the residual fraction in the IGM, then a
slightly different chain of events occurs. A larger initial fractional ionization implies that $t_{\rm rec}$ is shorter
and hence leads to a logarithmic increase in the amount of H$_{2}$ formed after a given physical time. The higher
H$_{2}$ fraction produced in this way allows the gas to cool to a slightly lower temperature, thereby boosting the
HD abundance to a significantly higher level than is reached in the low ionization case. This allows HD to take
over as the dominant coolant, driving the temperature down further. A number of studies have examined this
HD-dominated regime \cite[see e.g.][]{nu02,no05,jb06,yokh07,mb08,kreck10}, and have shown that the minimum
temperature reached in this case can be as low as the CMB temperature, and that the higher critical density of
HD,  $n_{\rm crit, HD} \sim 10^{6} \: {\rm cm^{-3}}$, means that the gas does not reach the loitering phase until
much later in its collapse. Once the gas does reach this phase, however, its subsequent evolution is very
similar to that in the low-ionization case discussed above. Cold gas accumulates at $n \sim n_{\rm crit}$
until its mass exceeds the Bonnor-Ebert mass, which in this case is $M_{\rm BE} \sim 40 \: {\rm M_{\odot}}$
if $T = 100$~K and $n = 10^{6} \: {\rm cm^{-3}}$. Once the gas mass exceeds $M_{\rm BE}$, the collapse
speeds up again, and the gas begins to heat up. Aside from the substantial difference in the size of $M_{\rm BE}$,
the main difference between the evolution of the gas in this case and in the low ionization case lies in the fact
that in the high ionization case, the gas reheats from $T \sim 200$~K or below to $T \sim 1000$~K much
more rapidly than in the low ionization case. As we shall see later, this period of rapid heating may exert an
important influence on the ability of the gas to fragment.

Several different scenarios have been identified that lead to an enhanced fractional ionization in the gas, and
that potentially allow the gas to reach the HD-dominated regime. Gas within minihalos with $T_{\rm vir} >
9000$~K will become hot enough for collisional ionization of hydrogen to supply the necessary electrons,
but halos of this size will typically have at least one star-forming progenitor \citep{jgb08,gg10}, and hence
typically will have already been enriched with metals. Other possibilities include the formation of
protogalaxies in ``fossil'' HII regions, i.e.\ regions that are recombining after having been ionized by
a previous Population III protostar \citep{oh03,no05,yokh07}, or that are irradiated with a significant
flux of X-rays \citep[e.g.][]{gb03} or cosmic rays \citep{sb07,jce07}. However, recent work by 
\citet{wgh11} has shown that the HD forming within a primordial minihalo is  very susceptible to photodissociation 
by even a weak extragalactic UV background \citep[see also][]{yoh07}, and so it is possible that gas will
reach the HD-dominated regime in only a few minihalos. 

\subsection{Three-body H$_{2}$ formation}
\label{threeb}
Once the collapsing gas reaches a density of around $10^{8}$--10$^{9} \: {\rm cm^{-3}}$, its chemical makeup starts 
to change significantly. The reason for this is that at these densities, the formation of H$_{2}$ via the three-body 
reactions \citep{pss83}
\begin{eqnarray}
{\rm H} + {\rm H} + {\rm H} & \rightarrow & {\rm H_{2}} + {\rm H}, \label{3b1} \\
{\rm H} + {\rm H} + {\rm H_{2}} & \rightarrow & {\rm H_{2}} + {\rm H_{2}}, \\
{\rm H} + {\rm H} + {\rm He} & \rightarrow & {\rm H_{2}} + {\rm He},
\end{eqnarray}
starts to become significant. These reactions quickly convert most of the hydrogen in the gas into H$_{2}$. At the same time,
however, they generate a substantial amount of thermal energy: every time an H$_{2}$ molecule forms via one of these
three-body reactions, its binding energy of 4.48~eV is converted into heat, and so at these densities, three-body H$_{2}$
formation heating can become the main process responsible for heating the gas. For this reason, even though the abundance 
of H$_{2}$, the dominant coolant during this phase of the collapse, increases by more than two orders of magnitude, the gas typically does not cool significantly. Indeed, the temperature often increases.

One major uncertainty in current treatments of the gas chemistry at this point in the collapse is exactly how quickly
the gas becomes molecular. Although reaction~\ref{3b1} is the dominant source of H$_{2}$ at these densities, the
rate coefficient for this reaction is poorly known, with published values differing by almost two orders of magnitude
at 1000~K, and by an even larger factor at lower temperatures \citep{g08,turk11}. The effects of this uncertainty have 
recently been studied by \citet{turk11}. They show that it has little effect on the density profile of the gas, and only a
limited effect on the temperature profile. However, it has far more significant effects on the morphology of the gas
and on its velocity structure. Simulations in which a high value was used for the three-body rate coefficient 
find that gravitational collapse occurs more rapidly, and show that the molecular gas develops a much more flattened, 
filamentary structure. Significant differences are also apparent in the infall velocities and the degree of rotational support.
The effect that this has on the ability of the gas to fragment (see Section~\ref{disk} below) is not yet understood.

\subsection{The final stages of collapse}
As the gas collapses to even higher densities, several important events occur which affect its ability to cool. The
first of these occurs at a density of around $10^{10} \: {\rm cm^{-3}}$, where the main rotational and vibrational lines of 
H$_{2}$ start to become optically thick. This reduces the
effectiveness of H$_{2}$ cooling,  leading to an ongoing rise in the gas temperature. In one-dimensional
simulations \citep[e.g.][]{on98,onus98,ripa02,ra04}, it is possible to treat optically thick H$_{2}$ cooling accurately by solving 
the full radiative transfer problem.  These models show that although the optical depth of the gas becomes large at frequencies 
corresponding to the centers of the main H$_{2}$ emission lines, the low continuum opacity of the gas allows photons
to continue to escape through the wings of the lines, with the result that the H$_{2}$ cooling rate is suppressed far
less rapidly as the collapse proceeds than one might at first expect (see \citealt{onus98} for a detailed discussion of this
point). In three-dimensional simulations, more approximate treatments of the cooling are necessary \citep[see e.g.][]{yoha06,tao09,clark11a},
but the general findings are the same -- H$_{2}$ cooling is suppressed to some extent, but still provides enough dissipation
of energy to allow the collapse to continue without a large increase in the temperature.

A second important event occurs once the number density reaches $n \sim 10^{14} \: {\rm cm^{-3}}$. At this density, 
collision-induced emission from H$_{2}$ (the inverse of the more familiar collision-induced absorption; see e.g.\
\citealt{fromm93}) becomes important. An isolated H$_{2}$ molecule has no dipole moment, but the activated complex
formed during the collision of two H$_{2}$ molecules can have a dipole moment, and hence can emit radiation via
dipole transitions. In principle, this can occur in gas of any density, but the 
probability of a photon being emitted in any given collision is very small, owing to the short lifetime of the collision 
state ($\Delta t < 10^{-12} \: {\rm s}$ at the temperatures relevant for Pop.\ III star formation; see \citealt{ra04}).
For this reason, collision-induced emission (CIE) becomes an important process only at very high gas densities. 
Another consequence of the short lifetime of the collision state is that the individual lines associated with the dipole 
transitions become so broadened that they actually merge into a continuum. This is important, as it means that the
high opacity of the gas in the rovibrational lines of H$_{2}$ does not significantly reduce the amount
of energy that can be radiated away by CIE. Therefore, once the gas reaches a sufficiently high density,
CIE becomes the dominant form of cooling, as pointed out by several authors \citep{on98,ripa02,ra04}.

The most detailed study of the effects of CIE cooling on the collapse of primordial gas was carried out
by \citet{ra04}. They showed that CIE cooling could actually become strong enough to trigger a thermal
instability, However, the growth rate of this instability is longer than the gravitational free-fall time, meaning
that it is unlikely that this process can drive fragmentation during the initial collapse of the gas. 

The phase of the collapse dominated by CIE cooling lasts for only a relatively short period of time. The
gas becomes optically thick in the continuum once it reaches a density  $ n \sim 10^{16} \: {\rm cm^{-3}}$
\citep{on98,ra04}, which strongly suppresses any further radiative cooling. Once this occurs, the gas
temperature rises until it reaches a point at which the H$_{2}$ begins to dissociate. At these densities,
this occurs at a temperature $T \sim 3000$~K. The dissociation of H$_{2}$ slows the temperature rise
for a while, as  most of the energy released during the collapse goes into dissociating the H$_{2}$ rather than 
raising the temperature. However, once the H$_{2}$ is exhausted, the temperature once again begins to
climb steeply. The thermal 
pressure in the interior of the collapsing core rises rapidly and eventually becomes strong enough to halt the 
collapse. At the point at which this occurs, the size of the dense core is around 0.1~AU, its mass is around
$0.01 \: {\rm M_{\odot}}$ and its mean density is of order $10^{20} \: {\rm cm^{-3}}$ \citep{yoh08}. It is bounded 
by a strong accretion shock. This pressure-supported, shock-bounded core is the structure that we identify as
a Population III protostar.

\subsection{Formation and fragmentation of the protostellar accretion disk}
\label{disk}
Most numerical simulations of the formation of Population III stars stop at or before the point at which 
the protostar forms, owing to the increasingly small timesteps that need to be taken to follow the
evolution of the gas. As significant fragmentation of the gas during the initial collapse is uncommon
(see e.g.\ \citealt{tao09}, who find that fragmentation into a binary system occurs in only one out of
five cases), most studies have assumed that the gas surrounding the protostar does not fragment 
at later times either, but instead is simply accreted by the existing protostar, either directly or via a 
protostellar accretion disk. A good overview of the results of these studies can be found in 
\citet{by11}.

However, in the past few years, a number of simulations have been performed that attempt
to directly model the evolution of the gas after the formation of the first Population III protostar
using a technique developed for studies of contemporary star formation \citep[see e.g.][]{cgk08,
sgb10,clark11a,clark11b,greif11,smith11}. Gravitationally bound 
regions of gas that become smaller than some pre-selected size scale are replaced by what are 
usually termed sink particles (in an SPH code; see e.g.\ \citealt{bbp95}) or sink cells (in a grid code; 
see e.g.\ \citealt{fed10}). These sinks can accrete gas from their surroundings and continue to interact 
gravitationally with the surrounding gas, but allow one to neglect the very small-scale hydrodynamical
flows that would otherwise force one to take very small numerical timesteps owing to the Courant 
condition. 

One of the main results of these simulations is the demonstration that the first protostar to form
does not simply accrete all of the available gas. The infalling gas forms an accretion disk around
the protostar, but this disk is unable to transfer gas inwards, onto the protostar, rapidly enough to
balance the rate at which new material is falling onto the disk. The disk therefore grows in mass
until it becomes gravitationally unstable, at which point it begins to fragment, forming multiple
protostars \citep{clark11b}. The gravitational interaction between these protostars and the surrounding
gas can trigger further fragmentation, and close encounters between protostars may lead to mergers,
or in some cases to protostars being ejected from the center of the minihalo \citep{greif11}.

Molecular hydrogen plays a crucial role in this process. The midplane temperature of the
protostellar accretion disk is typically 1500--2000~K, and the disk is maintained at this temperature 
primarily by a combination of H$_{2}$ line cooling and CIE cooling \citep{clark11b}. If the
molecular hydrogen were not present,  the main disk coolant would be bound-free emission
from H$^{-}$, which becomes effective only at a temperature $T \sim 6000$~K. The disk
would therefore be considerably hotter, allowing it to transfer gas more rapidly onto the 
central protostar, and making it less susceptible to gravitational instability. In this case,
previous studies have shown that the disk would probably remain stable and would not
fragment \citep{tm04,tb04,md05}. 

As well as regulating the disk temperature, molecular hydrogen also provides the
cooling necessary to allow the fragments forming in the disk to survive as distinct, 
gravitationally-bound objects. Studies of the fragmentation of self-gravitating accretion
disks have shown that in order for fragments to survive, they must be able to cool within
roughly half of an orbital period \citep{gammie01}. Fragments that cannot cool this quickly
undergo a thermal bounce and are then sheared apart by disk rotation, rather than continuing
to collapse. In the case of the fragments forming in a Pop.\ III accretion disk, two key processes
provide the necessary cooling: CIE cooling at $n < 10^{16} \: {\rm cm^{-3}}$ and the collisional
dissociation of H$_{2}$ at $n > 10^{16} \: {\rm cm^{-3}}$ \citep{clark11b}.

Finally, there are some indications that the chemical state of the gas at early times during its collapse
exerts a substantial influence on the degree of fragmentation that occurs at much later times. As we
have already seen, if HD cooling becomes efficient during the collapse of the gas, then the gas
can cool to a significantly lower temperature than in the standard H$_{2}$-dominated case. However,
a consequence of this is that the gas undergoes a more rapid period of reheating once the HD
molecules reach LTE, during which it has a very stiff effective equation of state. This leads to a
pronounced loss of small-scale structure from the gas, which appears to reduce the amount of
fragmentation that occurs at later times in the evolution of the minihalo \citep{clark11a,greif11}.


\section*{Acknowledgments}
Financial support for this work was provided by the
Baden-W\"urttemberg-Stiftung via their program International Collaboration II (grant P-LS-SPII/18), from the German
Bundesministerium f\"ur Bildung und Forschung via the ASTRONET project STAR FORMAT (grant 05A09VHA), and
by a Frontier grant of Heidelberg University sponsored by the German Excellence Initiative.

\bibliographystyle{apj}

\end{document}